\documentclass[acmtog]{acmart}

\usepackage{epigraph}
\usepackage{algorithm}
\usepackage[noend]{algpseudocode}

\usepackage{soul}
\usepackage{pifont}
\usepackage{xspace}
\usepackage[capitalise]{cleveref}
\usepackage{multirow}
\usepackage{wrapfig}
\usepackage[nolist]{acronym}
\usepackage{makecell}

\usepackage{siunitx}
\DeclareSIUnit{\sps}{smp/s}

\newcommand{\cmark}{\checkmark}%
\newcommand{\xmark}{\color{lightgray}\scalebox{0.85}{\ding{53}}}%

\DeclareGraphicsExtensions{.png,.jpg,.pdf,.ai,.psd}

\def\figurePath{figures/}

\newcommand{\eg}{e.g.,\xspace}
\newcommand{\ie}{i.e.,\xspace}

\newcommand{\refFig}[1]{Fig.~\ref{fig:#1}}
\newcommand{\refTab}[1]{Tab.~\ref{tab:#1}}
\newcommand{\refSec}[1]{Sec.~\ref{sec:#1}}
\newcommand{\refEq}[1]{Eq.~\ref{eq:#1}}
\newcommand{\refAlg}[1]{Alg.~\ref{alg:#1}}

\usepackage[bold=0.55]{xfakebold}
\newcommand{\winner}[1]{\setBold #1\unsetBold}

\usepackage{icomma}

\usepackage{xcolor}
\newcommand{\revision}[1]{#1}

\newcommand{\myfigure}[2]{%
    \begin{figure}[htb]%
    \centering\includegraphics*[width = \linewidth]{\figurePath#1}%
    \caption{#2}%
    \label{fig:#1}%
    \end{figure}%
}

\newcommand{\mycfigure}[2]{%
    \begin{figure*}[htb]%
    \centering\includegraphics*[width = \linewidth]{\figurePath#1}%
    \caption{#2}%
    \label{fig:#1}%
    \end{figure*}%
}

\newcommand{\mywfigure}[3]{%
\begin{wrapfigure}{r}{#2\columnwidth}%
  \begin{center}%
    \includegraphics[width=#2\columnwidth]{\figurePath#1}%
    \caption{#3}%
    \label{fig:#1}%
  \end{center}%
\end{wrapfigure}%
}

\soulregister\ref7
\soulregister\cite7
\soulregister\refFig7
\soulregister\refAlg7
\soulregister\cite7
\soulregister\ref7
\soulregister\pageref7
\soulregister\shortcite7
\soulregister\eg0
\soulregister\ie0
\soulregister\etal0

\newcommand{\mysection}[2]{\section{#1}\label{sec:#2}}
\newcommand{\mysubsection}[2]{\subsection{#1}\label{sec:#2}}
\newcommand{\mysubsubsection}[2]{\subsubsection{#1}\label{sec:#2}}

\definecolor{colorA}{HTML}{4285f4}
\definecolor{colorB}{HTML}{ea4335}
\definecolor{colorC}{HTML}{fbbc04}
\definecolor{colorD}{HTML}{34a853}
\definecolor{colorE}{HTML}{ff6d01}
\definecolor{colorF}{HTML}{46bdc6}
\definecolor{colorG}{HTML}{000000}
\definecolor{colorH}{HTML}{777777}
\definecolor{colorI}{HTML}{bdd6ff}
\definecolor{colorJ}{HTML}{6a9e6f}

\newcommand{\svn}[1]{\multicolumn1c{\small\texttt{\##1}}}

\begin{acronym}
\acro{BRDF}{bi-directional reflectance distribution function}
\acro{BVH}{bounding-volume hierarchy}
\acro{IGS}{International GNSS Service}
\acro{GPS}{global positioning system}
\acro{MC}{Monte Carlo}
\acro{MLP}{multi-layer perceptron}
\acro{ODE}{ordinary differential equation}
\acro{RMS}{root-mean square error}
\acro{SRP}{solar radiation pressure}
\acro{MEO}{medium earth orbit}
\acro{LEO}{low earth orbit}
\end{acronym}

\usepackage{bm}

\newcommand{\mymath}[2]{
    \newcommand{#1}{\TextOrMath{$#2$\xspace}{#2}}
}

\begin{document}

\setcopyright{cc}
\setcctype{by}
\acmJournal{TOG}
\acmYear{2026}
\acmVolume{45}
\acmNumber{4}
\acmArticle{82}
\acmMonth{7}
\acmDOI{10.1145/3811396}



\title{Photons$\,\times\,$Force: Differentiable Radiation Pressure Modeling}

\author{Charles Constant}
\affiliation{%
	\institution{University College London}
    \city{London}
	\country{United Kingdom}
}
\email{charles.constant.18@ucl.ac.uk}

\author{Elizabeth Bates}
\affiliation{%
	\institution{The Alan Turing Institute}
	\city{London}
    \country{United Kingdom}
}
\email{ebates@turing.ac.uk}

\author{Santosh Bhattarai}
\affiliation{%
	\institution{University College London}
	\city{London}
    \country{United Kingdom}
}
\email{s.bhattarai@ucl.ac.uk}

\author{Marek Ziebart}
\affiliation{%
	\institution{University College London}
	\city{London}
    \country{United Kingdom}
}
\email{m.ziebart@ucl.ac.uk}

\author{Tobias Ritschel}
\affiliation{%
	\institution{University College London}
	\city{London}
    \country{United Kingdom}
}
\email{t.ritschel@ucl.ac.uk}
\renewcommand{\shortauthors}{Constant et al.}

\begin{teaserfigure}  
    \includegraphics[width=\textwidth]{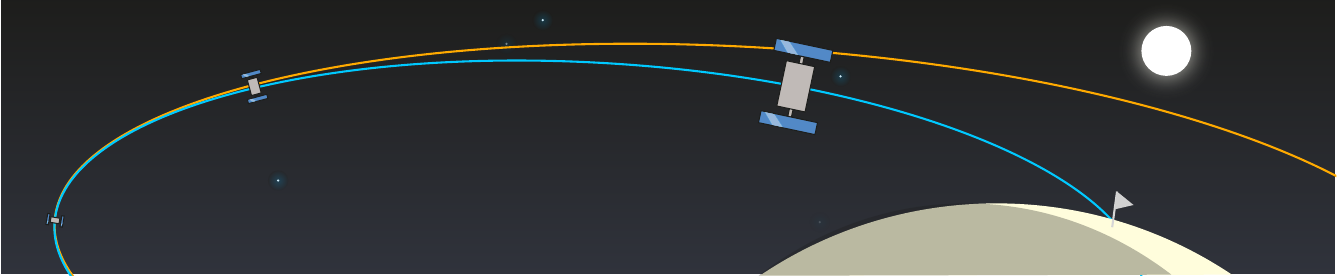}
    \caption{
    We devise a system for the automatic design of spacecraft (satellite, left) such that they arrive at a desired target (flag, right) along the blue path in the presence of radiation pressure (sun, top).
    Non-optimized systems or those not accounting for a combination of light and motion miss the target (orange path).}
    \label{fig:Teaser}
\end{teaserfigure}

\begin{abstract}
We propose a system to optimize parametric designs subject to radiation pressure, \ie the effect of light on the motion of objects.
This is most relevant in the design of spacecraft, where radiation pressure presents the dominant non-conservative forcing mechanism, which is the case beyond approximately 800 km altitude. Despite its importance, the high computational cost of high-fidelity radiation pressure modeling has limited its use in large-scale spacecraft design, optimization, and space situational awareness applications.
We enable this by offering three innovations in the simulation, in representation and in optimization:
First, a practical computer graphics-inspired Monte-Carlo (MC) simulation of radiation pressure.
The simulation is highly parallel, uses importance sampling and next-event estimation to reduce variance and allows simulating an entire family of designs instead of a single spacecraft as in previous work.
Second, we introduce neural networks as a representation of forces from design parameters.
This neural proxy model, learned from simulations, is inherently differentiable and can query forces orders of magnitude faster than a full MC simulation.
Third, and finally, we demonstrate optimizing inverse radiation pressure designs, such as finding geometry, material or operation parameters that minimizes travel time, maximizes proximity given a desired end-point, minimize thruster fuel, trains mission control policies or allocated compute budget in extraterrestrial compute.
\end{abstract}

\maketitle

\setlength\epigraphwidth{7.0cm}
\epigraph{"Once the rockets are up, who cares where they come down?
That's not my department!" says Wernher von Braun.}{Wernher von Braun, \textit{Tom Lehrer}}

\mysection{Introduction}{Introduction}
The understanding of celestial motion has always been an important driving force for the advancement of mathematics: from Stonehenge \cite{hawkins1963stonehenge}, to Galileo or Newton \cite{hall2012galileo}, the Apollo mission \cite{wollenhaupt1970apollo} or modern space monitoring \cite{enge1994global}, all relied on the most advanced mathematical methods and technology of their time.
Today, modern AI and GPUs are among the most advanced incarnation of mathematical technology, yet their full joint potential remains under explored in this context.

In this article we explore using AI and GPUs to advance the particular problem of predicting orbital trajectories as well as automatically designing spacecraft or their behavior to follow a desired trajectory (\refFig{Teaser}).
At present the accurate simulation of radiation forces on spacecraft either takes long (multiple days \cite{bhattarai2019demonstrating}) and/or is not accurate or compatible with modern AI, as it is not differentiable.
Hence the design of spacecraft is done by humans, not machines.
Yet, the understanding and control of such processes is vital to satellite-based navigation systems \eg GPS \cite{enge1994global}.

To unlock these abilities two conditions must be met: First, we need models that can be efficiently executed on GPUs at scale whilst remaining general and accurate enough \ie do not make too many assumptions.
Second, these models need to be differentiable, so as to be used in the modern optimization setting that underpins most of modern AI.
The difficulties in differentiation are two-fold: once in theory, as the integrand is discontinuous in visibility, and second, in practice, as existing differentiable rendering methods from computer graphics \cite{loper2014opendr,  li2018differentiable,loubet2019reparameterizing,vicini2021path} are both too slow to be used in a simulation loop and not general enough to deal with multiple designs.

The phenomenon we would like to model is \emph{solar radiation pressure} \cite{lebedev1901experimental}, the effect of radiation on motion: photons hitting an object impart momentum.
On Earth, in the presence of strong gravity, these effects are negligible. 
In outer space, for long trajectories and under strong radiation, these soon become relevant.
Often, this effect is a nuisance that presents challenges for the accuracy of orbit prediction \cite{rodriguez2014impact,montenbruck2000satellite} but has also been exploited to ease certain maneuvers \eg on the way to Mercury \cite{oshaughnessy2014messenger} or for general solar sailing \cite{johnson2011status}.
Ideally, the AI should not only predict and compensate for radiation pressure, but also exploit it to enhance or unlock new system capabilities. For example, to reduce fuel consumption or enable propellant-free maneuvers.

Links have been established in the literature between radiation pressure modeling and computer graphics when modeling solar radiation pressure \cite{ziebart2005combined}.
These methods are often limited to a few bounces, simplistic material models, or slow computation and they typically focus on a single specific design.
In contrast, we simulate an entire design space, and unlike current approaches, support optimization by providing differentiation.

Our contributions include:
\begin{itemize}
    \item A parallel and scalable method simulating radiation pressure with multiple bounces and general material properties
    \item Sampling of radiation pressure for entire design spaces
    \item Differentiation of said simulation through neural proxies.
    \item Inverse parametric spacecraft geometry, material, mission and control policy design using our neural proxy.
\end{itemize}

The resulting system is more accurate and faster than existing methods, allowing for rapid simulation of forces in a matter of seconds.
The use of a neural proxy makes this simulation an order of magnitude faster and, additionally, differentiable such that it can be used in an optimization context that requires repeated executions.
This unlocks the potential of automated satellite design in the presence of --but not limited to-- radiation pressure.
We demonstrate this by changing materials or geometry designs, so that a spacecraft hits a desired target at a prescribed time given initial conditions.

\mysection{Previous work}{PreviousWork}
The background to our work is threefold:
radiation pressure, an aspect of space science (\refSec{SolarPressureBG}) as well as   
inverse physics (\refSec{InversePhysicsBG}) and 
inverse rendering (\refSec{InverseRenderingBG}), both problems addressed in the computer graphics community. 

For a general background on satellite orbits and force modeling, refer to the textbook by \citet{montenbruck2000satellite}. For an in-depth guide to rendering, see PBRT by \citet{pharr2023physically}.

\mysubsection{Radiation Pressure}{SolarPressureBG}

Radiation pressure arises from the transfer of momentum carried by photons to an object \cite{lebedev1901experimental}: when a spacecraft in the vacuum of space is exposed to a photon flux, it experiences acceleration due to the cumulative momentum imparted by incident photons.
In particular, \ac{SRP}, can make a typical \ac{GPS} satellite drift by hundreds of meters a day \cite{springer1999new,montenbruck2015gnss,ziebart2005combined}.

The simplest and most common method to model the effect of solar radiation pressure starts with the so-called "cannonball" model \cite{montenbruck2015gnss}, which treats the spacecraft as a sphere characterized by certain reflectivity coefficients and a given area-to-mass ratio.
This approach was extended to the "box-wing" model, where radiation effects on each individual surface are analytically determined and summed.
Building on this, a frequently used approach in high precision applications augments the box-wing method with empirically derived coefficients, adding geometric terms to the accelerations and scaling them accordingly to compensate for deficiencies in the force model \cite{rodriguez2012adjustable}. While these techniques do improve orbit prediction accuracy, they reach a plateau: linking these back to physical details of the spacecraft is challenging, and thus provides few avenues for further enhancement of the understanding of the spacecraft's physical behaviour.

By contrast, maintaining a physics-based perspective allows continual refinement. We propose employing advanced computer graphics methods, notably ray tracing, to replicate the actual physical dynamics of the system, as detailed in \cite{ziebart2005combined,rodriguez2014impact}. This approach not only yields improved orbits but also preserves the ability to learn more about the spacecraft’s design and behavior over time.

\citet{li2018fast} and \citet{kenneally2019faster,kenneally2020fast} use a \ac{BVH} to accelerate ray-tracing based computation of solar radiation pressure.
This direct application is not possible to our problem, as the variation in design changes the geometry and hence a rebuild of a \ac{BVH} would be too slow to provide an advantage.
Also more or less advanced variants of refitting \cite{wald2007ray} the \ac{BVH} seem difficult, as changes are not one-dimensionally dependent on time and the geometry changes for every sample with no re-use to amortize over.
We provide experimental evidence in \refSec{Simulation} that na\"ive parallelization across samples and primitives is faster than BVH construction or refitting for our problem.

At GPS altitudes, direct solar radiation forces are overwhelmingly the contribution of SRP, but effects such as Earth radiation pressure \cite{rodriguez2012impact} are frequently accounted for in high-precision applications.
Our approach has the potential to accommodate radiation from multiple sources, including the Moon and even other planets for deep space missions. However, this capability has not yet been explored. The primary challenge in extending our model lies in implementing a differentiable radiative model for these bodies. In the case of Earth, such a model could be based on the CERES dataset \cite{wielicki1996clouds}.

Often in radiation pressure modelling, the geometry and materials of spacecraft are simplified.
For example, a decomposition into ``box'' and ``wing'' is made \cite{montenbruck2015gnss}, and where radiation force is modeled using ray-tracing, the computational cost is such that the solar arrays must be treated separately from the bus (satellite's main body holding the instruments).
This means shadowing and reflections between these two components are not accounted for \cite{ziebart2005combined}.
Our approach samples and represents the acceleration under time-changing geometry as a neural model, scaling to all kinds of variations in design or configurations of one design.

Closest in purpose to our approach is the use of \ac{SRP} as a means to plan and improve a mission to Mercury  \cite{oshaughnessy2014messenger}.
Their design not only considers \ac{SRP} as a disturbance but strategically leverages it to reduce steering effort by planning the trajectory in a way that utilizes \ac{SRP} to save fuel.
Our system enables the automation of such designs by differentiating the effects of \ac{SRP} to determine optimal spacecraft design parameters.

So far, the use of AI in spacecraft design has been limited to prediction or estimation of orbits \cite{caldas2024machine}, and did not account for \ac{SRP}, and even less for inverse design using it, such as we demonstrate with this work.

\mysubsection{Inverse physics}{InversePhysicsBG}
Our challenge is an instance of inverse differential equation solving \cite{isakov2006inverse} where observations and a differential equation are given and we estimate initial conditions.
Our particular problem is that we can only approximate using \ac{MC}; the integrand is not differentiable; and we explore many equations across many designs.
Earlier in graphics, the adjoint methods which we now used in an AI context, were previously developed in the computer graphics community for the control of smoke and fluid simulations \cite{mcnamara2004fluid}.
With the advent of AI, renewed interest has led to differentiable physics systems \cite{holl2020learning,toussaint2018differentiable,geilinger2020add,hu2019difftaichi,deavilla2018end}, but to our knowledge none have yet been applied to radiation, or more specific, to orbit force modeling yet.

The design of objects with a desired behavioral response to some physical phenomena has a tradition in graphics, where it was used to make object stand upright \cite{prevost2013make} or spin under forces \cite{bacher2014spin}.
More recently, \citet{allen2022physical} use an approach most similar to ours: they combine an adjoint solver with a surrogate model of fluid physics to design objects that provide desired behavior such as obstacles to water that splash desired targets.
We pursue a similar idea, but for the design of spacecraft under radiation pressure.

\mysubsection{Inverse Rendering}{InverseRenderingBG}
As explained in \refSec{Simulation}, \ac{SRP} is a non-trivial extension to rendering that involves several extra steps on top of determining incoming radiance which convert radiance to force.

Of particular interest in this paper is differentiating this operation, an ongoing effort in graphics named inverse rendering and explored by \citet{loper2014opendr}, \citet{li2018differentiable}, \citet{loubet2019reparameterizing}, \citet{vicini2021path}, or \citet{fischer2023plateau}.

In particular, here we treat the \ac{SRP} as a black box, for which we build a neural surrogate, or proxy model \cite{grzeszczuk1998neuroanimator,fischer2023zero} to make it differentiable.
We also implement a faster and more accurate model of \ac{SRP} than other solutions before, by making use of adequate generalization and implementation choices.
Finally, it is more general---both compared to classic \ac{SRP} solvers and to classic differentiable rendering--- as it deals with many varying designs and configurations at the same time so as to later optimize across them.

\mycfigure{Concept}{Main flow of our system: The input is a design space of spacecraft, shown here with different panel configurations as well as a design goal symbolized as a star.
In the first step, we simulate and store the force resulting from many directions of illumination under many different designs.
In the second step, a neural model of light direction and design parameters is learned which predicts force.
The final part is an optimization step that computes orbits of the given spacecraft (shown as spirals) using the forces of the neural model under a certain radiation, adjusting the design parameters so as to arrive at the goal (star).
The output are these optimal design parameters.
}
\mysection{Our approach}{OurApproach}

\mysubsection{Overview}{Overview}
Our system has three main parts (\refFig{Concept}): simulation (\refSec{Simulation}), representation (\refSec{Representation}) and optimization (\refSec{Optimization}).
Our system takes in two inputs: the \emph{design} and the \emph{problem}.
The outputs are the optimal design parameters to solve the problem.

The spacecraft design is formalized as a set of parameters, \eg the length of the solar arrays and the reflectance of the bus for a two-dimensional design space (\refFig{Concept}).
For simplicity, we will refer to the system simulated, represented and solved as a ``spacecraft'' but the approach could be applied to other systems as well.
We will also subsume all non-design parameters that affect the spacecraft into this design vector.
This removes the need to partition the model into separate components that do not interchange radiation in modeling, but would in reality, \eg panel orientation for  attitude laws.

We describe the \emph{simulation} of forces resulting from radiation for a particular system in \refSec{Simulation}.
\revision{This simulation is more accurate than previous approaches and often faster in comparison, though still too slow on its own for use inside an optimization loop.}
It is also more abstract than previous work, as it does not only represent radiation pressure for one design, but allows random access to any combination of design and radiation direction.
For our example above, this forms a 2+2-dimensional domain: two dimensions for the direction of radiation and two design space dimensions.
The simulation produces samples of a function mapping this domain to the range of 3D force and torque vectors.

In \refSec{Representation} we suggest how to \emph{represent} this complex function using a neural network. 
This network can be efficiently learned from batches of light directions, forces, torques and design parameters, resulting from simulations with random designs and random incoming radiation directions.
\revision{A neural proxy is faster than the simulation, differentiable, and noise-free; the last is essential as ODE integration with noisy MC force estimates is unstable (Sec. 4.3), and it sidesteps the interpolation schemes used by look-up-table approaches in spacecraft force modelling \cite{bhattarai2019demonstrating}.}
Note that the proxy is independent of the problem to solve, and depends only on the spacecraft geometry and design space.

The last part is the \emph{optimization} (\refSec{Optimization}).
Input to the optimization is the neural proxy trained in the previous steps and a problem description.
Optimization acts on the integration of spacecraft state over time in the presence of radiation.
This integration clearly depends on the design parameters: changing the color or geometry will change solar pressure, which will alter the trajectory and ultimately where the spacecraft will arrive, at what time and how fast.
A user specifies these desired target values, and the system optimizes the design to achieve these as closely as possible

We will now describe all three components in detail.

\mysubsection{Radiation Pressure Simulation}{Simulation}

\mymath{\lightDirection}{\bm\upomega}
\mymath{\localLightDirection}{\bar{\bm\upomega}}
\mymath{\forceField}{f}
\mymath{\force}{F}
\mymath{\torque}{\tau}
\mymath{\centerOfMass}{\location_\mathrm{COM}}
\mymath{\surface}{\mathcal M}
\mymath{\directions}{\Omega}
\mymath{\diff}{\mathrm d}
\mymath{\location}{\mathbf x}
\mymath{\incomdingDirection}{\lightDirection_\mathrm{in}}
\mymath{\outgoingDirection}{\lightDirection_\mathrm{out}}
\mymath{\normal}{\mathbf n}
\mymath{\brdf}{f_\mathrm r}
\mymath{\ltof}{\mathtt{l2f}}
\mymath{\radiation}{L}
\mymath{\speedOfLight}c
\mymath{\designParameters}{\theta}
\mymath{\proxyParameters}{\psi}
\mymath{\proxy}{P}
\mymath{\designDimension}{{n_\mathrm{d}}}
\mymath{\proxyDimension}{{n_\mathrm{p}}}
\mymath{\batchSize}{{n_\mathrm{s}}}
\mymath{\stepCount}{{n_\mathrm{n}}}
\mymath{\optimizationStepCount}{{n_\mathrm{o}}}
\mymath{\throughput}t
\mymath{\reflectance}r
\mymath{\geometricTerm}g
\mymath{\samples}{\xi}
\mymath{\referenceState}{\trajectory_\mathrm{Ref}}
\mymath{\proxyLearningRate}{\lambda_\mathrm{p}}
\mymath{\optimizationLearningRate}{\lambda_\mathrm{o}}

For the radiation simulation, we make the assumption that light is infinitely far away and comes from a single direction \lightDirection.
The acceleration imparted to the spacecraft is a function of this direction.

\myfigure{Rendering}{Radiation pressure converting light to force:
Radiation arrives from one direction \incomdingDirection and is reflected to another random direction \outgoingDirection at a position \location, with a probability that depends on the material and the incoming direction.
The resulting force \force is the negated weighted half-vector of these.
}

The spatially-varying force field
$
\forceField(\location,\lightDirection,\designParameters)
\in
\mathbb R^3\times
\revision{\directions}\times
\mathbb R^\designDimension
\rightarrow
\mathbb R^3$
resulting from \revision{illumination at a surface point $\location\in\mathbb R^3$ of a design with
design parameters}
$
\designParameters\in
\mathbb R^\designDimension$
by radiation from direction $
\lightDirection\in\directions$
is the integral
\begin{align}
\label{eq:Force}
\forceField(\location,\lightDirection,\designParameters)
=
\int_
{\directions\times\directions}
\ltof
\left(
\radiation(
\location,
\incomdingDirection,
\designParameters),
\incomdingDirection,\outgoingDirection
\right)
\diff\incomdingDirection
\diff\outgoingDirection
\end{align}
where
\radiation is radiation from \incomdingDirection and \ltof is the light-to-force-mapping
\begin{align}
\ltof(\radiation,\incomdingDirection,\outgoingDirection)
=
\radiation
\left<\normal,\incomdingDirection\right>^+
\left(
-\incomdingDirection
-
\outgoingDirection
\brdf(\incomdingDirection,\outgoingDirection)
\right)
\speedOfLight^{-1}
,
\end{align}
with \speedOfLight being the speed of light and $<>^+$ the clamped dot product \cite{montenbruck2000satellite}\revision{, and \brdf the \ac{BRDF}}.
\revision{Note that \lightDirection is a scene parameter, the sun direction, rather than an integration variable.}

The corresponding translational force vector
$
\force(\lightDirection,\designParameters)
\in
\revision{\directions}\times
\mathbb R^\designDimension
\rightarrow
\mathbb R^3$
is defined by an additional integral over the surface
\begin{align}
&
\force(\lightDirection,\designParameters)
=
\int_
{\revision{\surface(\designParameters)}}
\forceField(\location,\lightDirection,\designParameters)
\diff\location
\end{align}

Similarly, the rotational force (torque) $
\torque(\lightDirection,\designParameters)
\in
\revision{\directions}\times
\mathbb R^\designDimension
\rightarrow
\mathbb R^3$
is computed as
\begin{align}
\torque(\lightDirection,\designParameters)
=
\int_
{\revision{\surface(\designParameters)}}
(\location-\revision{\centerOfMass(\designParameters)})
\times
\forceField(\location,\lightDirection,\designParameters)
\diff\location,
\end{align}
\revision{where \centerOfMass(\designParameters) is the spacecraft's center of mass, which like \surface(\designParameters) depends on the design.}

The integrals over the surface and all pairs of incoming and outgoing directions have an integrand with two parts: radiation and conversion of radiation into force.

The radiation component resembles a light field, which can be computed using computer graphics techniques \cite{gortler96lumigraph}.
This differs in so far that, it matters for force where the light has come from, while a classic light field only depends on position and outgoing direction ---its signature would be $\radiation(\location,\outgoingDirection)$--- while all the incoming directions were integrated over, weighted by the \ac{BRDF}.

Conversion of radiance into force involves the sum of both the direction of illumination and the direction of reflection.
For correctness of units, this is divided by the speed of light, \speedOfLight.

\paragraph{Estimation}
We solve for the incoming light of the integrand in \refEq{Force} using Monte Carlo path tracing \cite{veach1998robust,pharr2023physically}.
Note that the solution is not always found for a specific light direction \lightDirection and a specific design parameter vector \designParameters in one thread, but for many of these in a batch.

The spacecraft is given as a mapping from design parameters to triangle meshes with per-face Phong material parameters.
Note that this negates the usual advantage of \acp{BVH} \cite{macdonald1990heuristics,goldsmith1987automatic} or other forms of caching, which have been applied in previous \ac{SRP} work \cite{li2018fast} as each samples' ray intersects a different design, and hence a different geometry \ie different BVH.

To evaluate a single sample of the force, a random triangle on the surface is chosen by sampling according to triangle area and a random point within that triangle using uniform point picking \cite{weinstein2025triangle}.
Next, we pick a random incoming direction and evaluate the radiation using a single sample of path tracing.
Finally, a random outgoing direction is selected, and the force is calculated using the second part of the integrand in the equation \refEq{Force}.

We use \ac{BRDF} importance sampling to generate the incoming and outgoing direction.
We also use next-event estimation to obtain the incoming radiance \ie we follow the path as usual, but at every light path vertex we check direct light from \lightDirection.

\begin{equation}
     \force(\lightDirection, \designParameters) = \frac{A}{N} \sum_{i=1}^{N} \frac{\radiation_i}{c} \, \langle
  {\normal}_i, {\incomdingDirection}_i \rangle^{+} \, \frac{{\incomdingDirection}_i +
  \reflectance({\incomdingDirection}_i, {\outgoingDirection}_i) \,
  {\outgoingDirection}_i}{p({\outgoingDirection}_i)}
\end{equation}

where $A$ is the total surface area, $p({\outgoingDirection}_i)$ the \ac{BRDF}-importance-sampling PDF, and $\radiation_i$ the path-traced radiance with direct illumination from \lightDirection added at each vertex by next-event estimation \cite{veach1998robust}.
This is a surrogate: we sample to reduce variance of \radiation, but the quantity of interest is \force, related to \radiation through \ltof non-trivially. Force-side importance sampling is left as future work.

Note that due to the random designs we sample, each ray effectively intersects a different world.
Thus, creating one unique \ac{BVH} per unique design per ray is impractical and instead, we resort to a basic, but massively parallel intersection test.
This is done in batches in the orders of ten-thousands of rays.
Then each of these rays, for each bounce, is tested against all triangles in parallel.
A typical mesh has in the order of thousands of polygons, resulting in millions of threads, which produce a reasonable spread of work on modern compute architectures.
Devising a form of hyper-BVH for arbitrary design parameters (such as done for special cases of \eg motion blur \cite{woop2017stbvh}) to perform efficient multi-world ray-tracing remains as future work.
The system is implemented in JAX \cite{jax2018github} using \texttt{for\_i} and \texttt{scan} primitives from XLA \cite{sabne2020xla} to avoid Python loops.

Note that the output of this step is a list of pairs consisting of light directions and design parameters mapped to force and torque, disregarding surface locations.
Subsequently, and after integration over the surface, the spacecraft is treated as a single point, responding to directional light with directional force and torque.

\revision{Sampling the design space is also non-trivial: at training time we estimate gradients of proxy parameters with respect to a force loss, not the force itself, and efficient estimation of those is future work.}

\mysubsection{Representation}{Representation}
Running the simulation derived in the previous section, we can in principle query the force \force corresponding to a spacecraft for any design parameter \designParameters and any radiation direction \lightDirection.
In practice, this has two issues if used inside an optimization loop which is again a loop over many steps of solving for the spacecraft's trajectory.
First, it takes too long to compute a force sample by simulating radiation and second, it cannot be differentiated (in theory due to the discontinuity in visibility and if that is addressed, not with practical computational efficiency).
To overcome both issues, we make use of neural proxies \cite{fischer2023zero,grzeszczuk1998neuroanimator}.
A neural proxy here is a neural network that takes as input the design parameters and the light direction, and returns 3D force and torque vectors.
It is parametrized by $\proxyParameters\in\mathbb R^\proxyDimension$, another set of parameters (not to be confused with the design parameters, \designParameters).
The proxy \proxy has the same input and output as the force simulation it emulates.
It is supervised on an infinite stream of random samples of designs under random radiation directions, \ie minimizing the empirical risk
\begin{equation}
\operatorname{argmin}_\proxyParameters 
\mathbb E_{\lightDirection,\designParameters}
\left[
||
\proxy(\lightDirection,\designParameters|\proxyParameters)-
\left(
\force(\lightDirection,\designParameters)|
\torque(\lightDirection,\designParameters)
\right)
||
\right]
.
\end{equation}

\paragraph{Architecture}
In detail, the proxy is represented using a \ac{MLP} with three layers, and 16 hidden states, followed by ReLUs and a final sigmoid (shifted to produce outputs from -1 to 1), which is appropriate as the force and torque resulting from unit radiation is bound.
We experimented with (directional) positional encoding, but did not observe substantial benefits.
We implement the \ac{MLP} in JAX and train it using ADAM, which typically takes less than a minute on a T4 GPU.
We train on reasonably-converged but far from noise-free \ac{MC} samples, assuming the network will act as a denoiser \cite{vincent2008extracting,lehtinen1803noise2noise}.
All input parameters are bound, so the network never needs to extrapolate: the light direction is in the angular domain, and design parameters are clipped in the optimization anyway, \ie never required outside the range they were trained on.

\mymath{\trajectory}{\bm z}
\mymath{\positionTrajectory}{\trajectory^\mathrm{Pos}} 
\mymath{\velocityTrajectory}{\trajectory^\mathrm{Vel}}
\mymath{\orientationTrajectory}{\trajectory^\mathrm{Ori}} 
\mymath{\angularVelocityTrajectory}{\trajectory^\mathrm{Ang Vel}}
\mymath{\timeCoord}{t}
\mymath{\timeCoordStart}{{t_0}}
\mymath{\timeCoordEnd}{{t_1}}
\mymath{\loss}g
\mymath{\trajectorySize}m
\mymath{\optimalDesignparameters}{\designParameters^\star}
\mymath{\stateStart}{\trajectory(\timeCoordStart)}
\mymath{\stateEnd}{\trajectory(\timeCoordEnd)}
\mymath{\ode}{\dot\trajectory}
\mymath{\otherForces}{\bm o}
\mymath{\spaceshipMass}m
\mymath{\spaceshipMomentOfInertia}{\mathsf{I}}
\mymath{\localFrame}{\mathsf T}
\mymath{\weightPost}{\alpha_\mathrm{Post}}
\mymath{\weightFinal}{\alpha_N}
\mymath{\weightCol}{\alpha_\mathrm{Col}}
\mymath{\deviationPost}{\bar{d}_\mathrm{Post}}
\mymath{\deviationFinal}{d_N}
\mymath{\deviationCol}{\bar{d}_\mathrm{Col}}
\mymath{\kernelWidth}{\sigma}
\mymath{\quaternionRateMatrix}{\boldsymbol{\Omega}}
\mysubsection{Optimization}{Optimization}

Inputs for optimization are the force and torque proxy \proxy of a specific spacecraft design space, the light direction \lightDirection as well as a goal \loss to be defined next.
Output is the optimal set of parameters \optimalDesignparameters.

Optimization relates to the initial, simulated and final states of the spacecraft, which are thirteen-dimensional state vectors 
\begin{equation}
\trajectory=
(
\positionTrajectory, 
\velocityTrajectory,
\orientationTrajectory, 
\angularVelocityTrajectory
)
\end{equation}
 that stack 3D position, 3D velocity, quaternion orientation and angular velocity.
This goal is formulated as a loss that takes as input specific states (e.g. the end state \stateEnd) output by running an integrator from initial conditions \stateStart using the design to optimize.
To understand this relation, we will first recall how an integrator works and then define our loss.

\mysubsubsection{Integration}{Integration}

The integration step computes the state at time \timeCoord for a given design \designParameters and given initial conditions \stateStart:

\begin{equation}
\trajectory(\timeCoord)
=
\trajectory(\timeCoordStart)
+
\int_\timeCoordStart^\timeCoordEnd
\ode(\trajectory(\timeCoord),  \timeCoord, \designParameters)
\diff \timeCoord
\end{equation}

where \ode is the \ac{ODE} underlying a spacecraft's motion by velocity and acceleration:

\begin{equation}
\label{eq:ODE}
\ode(\trajectory, \timeCoord, \designParameters)=
\left(
\begin{array}{cc}
    \trajectory^\mathrm{Vel}
    \\
    \spaceshipMass^{-1}
    \cdot \localFrame^\top \cdot \force(\localLightDirection, \designParameters) + 
    \otherForces(\trajectory)\\
    \frac{1}{2} \quaternionRateMatrix(\angularVelocityTrajectory) \, \orientationTrajectory
    \\
    \spaceshipMomentOfInertia^{-1}
    (
    \torque(\localLightDirection, \designParameters)-
    \angularVelocityTrajectory
    \times
    (\spaceshipMomentOfInertia\angularVelocityTrajectory)
    )
\end{array}
\right)
\end{equation}

The change of position is, trivially, the velocity, the first component.
The second component combines \revision{acceleration} due to SRP with other \revision{accelerations}, \otherForces, such as gravity or atmospheric drag, which can be included as long as they are differentiable.
Here, $\localLightDirection=\localFrame(\trajectory)\lightDirection$ is the local light direction, where $\localFrame$ is the rotation matrix corresponding to the quaternion orientation $\orientationTrajectory$.
As \force is represented in the spacecraft body-fixed reference frame it must be multiplied by $\localFrame^\top$ to convert the accelerations to inertial space \cite{montenbruck2015gnss}.
The third component is the change of orientation, given by the quaternion kinematic equation $\frac{1}{2}\quaternionRateMatrix(\angularVelocityTrajectory)\orientationTrajectory$, where \quaternionRateMatrix is a $4 \times 4$ matrix that maps angular velocity to quaternion derivative \cite{schaub2018}.
The fourth component is the angular acceleration which can be computed from \torque due to radiation pressure, the spacecraft's moments of inertia \spaceshipMomentOfInertia and the gyroscopic coupling term \cite{schaub2018}.
We do not consider other rotational forces.

As additional \revision{accelerations} \otherForces, we model the gravity field of Earth, and point mass accelerations due to the Sun and Moon, according to \citet{montenbruck2015gnss}.
In particular, for the Earth's gravity we use the EGM2008 model \cite{pavlis2012development} to degree and order 18, and make use of the JPL DE 441 planetary ephemerides \cite{park2021jpl} to query the position of the Sun and Moon.
A cylinder eclipse model is implemented to deal with the conditions under which the spacecraft enters solar and lunar eclipse \cite{montenbruck2015gnss}.
We do not model solid Earth tides, gravity from other bodies in the solar system or relativistic effects.

\paragraph{Implementation}
We use a typical fourth-order Runge-Kutta integrator \cite{wanner1996solving} with a 15-second step size to perform this task, implemented in JAX \cite{jax2018github}.
The integrator is again implemented as a XLA \texttt{for\_i} primitive to avoid loops \cite{sabne2020xla}.
Computation is done in double precision.
Integration is carried in an Earth-centered inertial reference frame (so-called J2000 reference frame \cite{montenbruck2015gnss}).

\mysubsubsection{Loss}{Loss}
The loss $\loss\in\mathbb R^{13}\rightarrow\mathbb R$ evaluates the state resulting from choice of design parameters in combination with some initial conditions.
The goal is expressed through a differentiable loss function that evaluates the end state, or in some cases also intermediate ones, and produces a positive value, which is smaller when the design is more desirable.
A simple example loss function would compare the last position of a trajectory to a reference 
$
\loss(\trajectory) = 
||
\positionTrajectory-
\positionTrajectory_\mathrm{Ref}
||
,
$
or ask a satellite to be stable
$
\loss(\trajectory) = 
||
\angularVelocityTrajectory
||
.
$
\mysubsubsection{Cost function}{Cost}
The final term to optimize for is

\begin{equation}
\optimalDesignparameters
=
\operatorname{argmin}_\designParameters
\loss(\trajectory(\timeCoordEnd,\designParameters))
\end{equation}

which involves simulating the trajectory from \timeCoordStart to \timeCoordEnd and running the loss on the resulting state.
Determining these parameters requires differentiating through the ODE solver.
This is enabled by using the adjoint method \cite{pontryagin1962mathematical} popularized in NeuralODE \cite{chen2018neural}.
As some design parameters only allow for a limited range (\eg\ reflectance), they get clipped after each optimization step.

\mysubsection{Pseudocode}{Pseudocode}
We will now discuss the full algorithm again, including batching, using the pseudo-code in \refAlg{Main}.
For conceptual simplicity, we explain the approach using infinitely sized-batches in this pseudo code.
In practice, batches are finite, and need to be aggregated by interleaving the simulation and the NN fitting.

\newcommand{\algmul}{$\times$\xspace}
  \begin{algorithm}[b]
      \caption{\label{alg:Main}Pseudo-code of our complete approach.
      Note that some functions are batched and vectorized, Python-style.}
      \begin{algorithmic}[1]
      \Procedure{Main}
      {\referenceState,\lightDirection, \surface, \batchSize, \designDimension, \stepCount, \proxyLearningRate, \optimizationStepCount, \optimizationLearningRate}
          \State \proxyParameters = \Call{Rand}{ }
          \label{line:LearningStart}
          \For{\stepCount}
              \State \proxyParameters = \proxyParameters - \proxyLearningRate \algmul \Call{\Call{Grad}{NnLoss}}{\proxyParameters}
          \EndFor
          \label{line:LearningEnd}
          \State \designParameters = \Call{RandDesign}{\designDimension}
          \label{line:OptimizationStart}
          \For{\optimizationStepCount}
              \State \designParameters = \designParameters - \optimizationLearningRate \algmul \Call{\Call{Adjoint}{ODE, ODELoss}}{\designParameters, \timeCoordStart, \timeCoordEnd}
          \EndFor
          \State\Return\designParameters
          \label{line:OptimizationEnd}
      \EndProcedure
      \Procedure{NNLoss}{\proxyParameters}
          \label{line:NNLoss}
          \State \designParameters = \Call{RandDesign}{\designDimension}
          \label{line:SimulationStart}
          \State \location = \Call{RandSurfPos}{\batchSize, \surface}
          \State \incomdingDirection = \Call{RandDir}{\batchSize}
          \State \outgoingDirection = \Call{RandDir}{\batchSize}
          \State \throughput = \Call{PathTrace}{\designParameters, \position, \incomdingDirection}
          \State \reflectance = \Call{BRDF}{\location, \incomdingDirection, \outgoingDirection}
          \State \geometricTerm = \Call{ClampedDot}{\normal, \incomdingDirection}
          \State \force = \throughput \algmul \geometricTerm \algmul (\incomdingDirection + \reflectance \algmul \outgoingDirection) / \speedOfLight
          \label{line:Force}
          \label{line:SimulationEnd}
          \State \Return |\proxy(\lightDirection, \designParameters, \proxyParameters) - \force|
      \EndProcedure
      \Procedure{ODELoss}{\trajectory}
          \label{line:ODELoss}
          \State\Return |\trajectory-\referenceState|
      \EndProcedure
      \Procedure{ode}{\trajectory, \designParameters, \timeCoord}:
          \State $\velocityTrajectory= \spaceshipMass^{-1} \cdot \force(\localLightDirection)$
          \State $\angularVelocityTrajectory= \spaceshipMomentOfInertia^{-1}( \torque(\localLightDirection, \designParameters) - \angularVelocityTrajectory \times ( \spaceshipMomentOfInertia \angularVelocityTrajectory )$
          \State \Return $(\positionTrajectory, \velocityTrajectory, \orientationTrajectory, \angularVelocityTrajectory)$
      \EndProcedure
      \end{algorithmic}
  \end{algorithm}

\newcommand{\refLine}[1]{\texttt{L\ref{line:#1}}}
In \refLine{SimulationStart} to \refLine{SimulationEnd} we perform the simulation, where we first sample \batchSize samples in one batch from their distribution using the appropriate functions.
We then call a path tracer, which also works on batches.
We have explained the parallelization above, but importantly, it returns the radiation for all samples in the batch.
We then compute \refEq{Force} in \refLine{Force}, which completes the simulation.
This process yields the resulting force, design parameters, and directions.

\refLine{LearningStart} to \refLine{LearningEnd} is the learning loop.
First we initialize the proxy network parameters \proxyParameters, before updating them with the gradient of the loss.
\textsc{Grad} is an operator that takes a function and returns a function to compute its gradient, a part of JAX.
The loss function in \refLine{NNLoss} \revision{$L_1$}-compares the simulated force to the one inferred by the network. \revision{We use the $L_1$ norm as it is more robust to outlier MC samples than $L_2$.}

In \refLine{OptimizationStart} the actual optimization starts by initializing a single design (not a batch) and adjusting it up to \refLine{OptimizationEnd}.
Similar to \textsc{Grad}, we compute the gradient using the operator \textsc{Adjoint} that takes as argument two functions: an ODE and a loss.
The loss in this case is seen in \refLine{ODELoss} to compare the end point of the trajectory to the desired one in $L_2$.
It then returns another function, that computes the gradient of the loss according to the integration of the ODE from start to finish.
Once the for loop concludes, the optimal parameters are determined and ready to be returned and deployed.

Overall, the code of our entire system is simple, comprising of four Colab notebooks (simulation, NN fitting, ODE solving, optimization) with a few hundred lines of code each with no external dependencies other than JAX and Optax.

Code is available at \url{https://github.com/CharlesPlusC/photonsXforce}.

\mysection{Evaluation}{Evaluation}
Given the immense practical difficulty of actually launching multiple spacecraft, we evaluate our approach in a cascade of simulations.
For this methodology to be valid, we first provide evidence that the simulation is in agreement with reality and in a second step use our system in these simulated conditions.

The logic of this is explained in \refFig{Dependency}: for our overall system to work, the optimization through the ODE must work.
This combines three aspects:
the neural proxy has to be representative of solar pressure in actual satellites geometries (\refSec{RepresentationVerification}),
the integrator has to be able to predict trajectories from given initial conditions (\refSec{IntegratorVerification}) and the adjoint solve has to be able to optimize ``through'' the ODE found in the prediction of spacecraft orbits (\refSec{OptimizationVerification}).
Then again, for the proxy to be representative of the solar pressure, the solar pressure itself has to be right (\refSec{SimulationVerification}).
To be ``right'' here never means perfect, but we require it to be on par with the state-of-the art and possess the other desired properties (\ie speed, differentiability, etc).
We consider two important components our method relies on, to be validated elsewhere: the forces  and the ability of an adjoint in general to invert ODEs \cite{chen2018neural}.

\myfigure{Dependency}{Evaluation dependency. Edges are \emph{works-if} relationships.
Nodes are evaluation steps.
Grey components are parts of the system, but we trust they were verified elsewhere.}

\newcommand{\method}[1]{\textsc{#1}}
An overview of all methods is shown in \refTab{Methods}.
We will progressively detail all aspects of a method as they become relevant in each step of the evaluation.

\begin{table}[!htp]
\setlength{\tabcolsep}{3.8pt}
\centering
\caption{Summary of all \ac{SRP} methods (rows) used in our evaluation in respect to different properties (columns).
Speed is in million \ac{SRP} samples per second on an Nvidia A100, where more is better.
\revision{Scalability is the exponent $\alpha$ in $t\propto n^\alpha$ ($t$ compute time, $n$ samples; $\alpha{=}1$ sequential, $\alpha{=}0$ parallel; less is better).}
``Bounce'' and ``BRDF'' are ticked if a method supports light bounces and BRDF, respectively.
``Differ'' is for a differentiable method.
``Grid'' and ``Neural'' are for the respective implementations.
The ticks in ``Evaluation'' indicate if a method is part of one aspect of our evaluation: integration, representation or optimization.}
\label{tab:Methods}
\begin{tabular}{lrrrcccccccc}
\toprule
&&&&&&&&&\multicolumn3c{Evaluation}\\
\cmidrule{10-12}
\multicolumn1c{Method}&
\multicolumn1c{Speed}&
\rotatebox{90}{Scaling}&
\rotatebox{90}{Bounce}&
\rotatebox{90}{BRDF}&
\rotatebox{90}{Design}&
\rotatebox{90}{Differ.}&
\rotatebox{90}{Grid}&
\rotatebox{90}{Neural}&
\rotatebox{90}{Integ.}&
\rotatebox{90}{Repres.}&
\rotatebox{90}{Optim.}
\\
\midrule
\method{Ball}&20B&1.0&\xmark&\xmark&\xmark&\xmark&\xmark&\xmark&\cmark&\xmark&\xmark\\
\method{ClassicGrid}&1M&0.5&\xmark&\xmark&\xmark&\xmark&\cmark&\xmark&\cmark&\xmark&\xmark\\
\method{RTGrid}&1M&0.5&\cmark&\cmark&\cmark&\xmark&\cmark&\xmark&\xmark&\xmark&\xmark\\
\method{BVHRTGrid}&1M&0.5&\cmark&\cmark&\xmark&\xmark&\cmark&\xmark&\xmark&\xmark&\xmark\\
\method{OursGrid}&10M&0.3&\xmark&\cmark&\cmark&\xmark&\cmark&\xmark&\cmark&\xmark&\xmark\\
\method{OursGridNB}&10M&0.2&\cmark&\cmark&\cmark&\xmark&\cmark&\xmark&\cmark&\xmark&\xmark\\
\method{OursNN}&500M&0.2&\cmark&\cmark&\cmark&\cmark&\xmark&\cmark&\cmark&\cmark&\cmark\\
\bottomrule
\end{tabular}
\end{table}

\mysubsection{Integration}{IntegratorVerification}

We first verify that our integrator can solve the relevant problems --motion of spacecraft in outer space-- with an error similar to a reference integrator on a dataset of representative tasks.

\paragraph{Dataset}
As tasks we use prediction of all 12 GPS block 2-F satellites on the 1st of March 2020.
12 trajectories are published by the \ac{IGS} \cite{johnston2017international}.
These represent the gold standard in the satellite positioning community.
These are routinely compared to laser ranging and have a reported accuracy on the scale of a few \unit{cm}  \cite{zajdel2023combination}.

\paragraph{Metric}
We simulate the trajectory for the first 12 hours and \qty{2880} time steps and compute \ac{RMS} between this trajectory and the baseline trajectory (less is better).

\paragraph{Baseline}
Our initial step is to benchmark our integrator against \method{Orekit}, a state-of-the-art library that "provides accurate and efficient low-level components for flight dynamics" \cite{maisonobe2010orekit}. We begin by initializing both our integrator and Orekit with identical initial conditions and force models (Earth gravity to degree and order 18 and luni-solar gravity).
If the two integrators yield consistent results under the same conditions, we can confirm that our integrator performs as expected.

\begin{wraptable}r{0.42\linewidth}
    \centering
    \caption{RMS error between the trajectories output by our integrator and Orekit in meters.
    }
    \label{tab:IntegrationVerification}
    \begin{tabular}{cr}
        SVN&
        \makecell{$\Delta$ RMS\\to \method{Orekit}}\\
        \toprule
\svn{63}&0.07\\
\svn{69}&0.12\\
\svn{67}&0.05\\
\svn{72}&0.09\\
\svn{68}&0.07\\
\svn{73}&0.07\\
\svn{65}&0.05\\
\svn{62}&0.27\\
\svn{71}&0.18\\
\svn{66}&0.07\\
\svn{64}&0.16\\
\svn{70}&0.05\\
\midrule
Mean&0.10\\
        \bottomrule
    \end{tabular}
\end{wraptable}

\paragraph{Results}
The results, shown in \refTab{IntegrationVerification}, indicate that our integrator achieves errors generally at the decimeter level or below, with the average error being at the decimeter level. To put this into context, the reported error occurs after the satellite has traveled approximately 170,000 kilometers over a 12-hour period. We believe that the observed discrepancies are potentially due to differences in the gravity models used—\method{orekit} employs EIGEN6S4, a different model compared to the EGM 2008 model that we use. Nevertheless, the overall error remains low in absolute terms.
While this does not yet provide an immediately practical advantage, it confirms that our integrator is state-of-the-art and possesses the essential property for our ultimate goal: full differentiability.

\mysubsection{Simulation}{SimulationVerification}
We will now compare our method for computing forces due to solar radiation pressure with other approaches listed in \refTab{Methods}.

\paragraph{Methods}
We compare the cannonball method (\method{Ball}), a well-known physics-based ray-tracing approach \cite{ziebart2005combined} which represents the state-of-the-art in physics-based radiation force modelling \cite{bhattarai2022high} (\method{ClassicGrid}), as well as our immediate method (\method{OursGrid}) and its neural proxy (\method{OursNN}).

The \method{Ball} method is simple and analytical, enabling the generation of a large number of samples, but it does not account for bounces or BRDFs.
The \method{ClassicGrid} method supports any number of bounces but is CPU-bound and does not support BRDF use. \revision{\method{RTGrid} is the ray-tracing approach of \citet{ziebart2005combined} with BRDF support, and \method{BVHRTGrid} is its BVH-accelerated variant \cite{li2018fast}. \method{OursGridNB} is our method without light bounces, included to isolate the contribution of indirect illumination.} Only our method supports design-dependent \ac{SRP}, but for this experiment, we consider a zero-dimensional design space. We will get back to evaluation of this soon.

\paragraph{Metrics}
As the main metric, we use the 3D RMS between the observed and the simulated trajectories (less is better).
The trajectories are computed by using each method as a force in the integrator which we have validated in the previous step, (\refSec{IntegratorVerification}).
As additional metrics, we evaluate bandwidth, measured in millions of samples per second (higher is better), and scalability, assessed by the slope of the function mapping the number of samples to compute time (lower is better).

\paragraph{Result}
Results are seen in \refTab{SimulationVerification}.
We find that our method has the lowest  error on average.
We outperform all baselines in 10 of all 12 tracks. 
While our main aim is adding differentiability and generalization over designs, this non-inferiority is still a necessary condition for success.

Please refer to the column ``Speed'' in \refTab{Methods} for computer times.
We see that our method performs order of magnitudes faster.
For \method{RTGrid} we could not execute the code for comparison on an Nvidia A100, but calculated the number of samples per second from the total compute time and number of rays stated for a 100-CPU cluster. 

Additionally, the column ``Scalability'' indicates that our method also scales better to more subtle \ac{SRP} as it grows less in compute demand as the number of samples increases.

\begin{table*}[htb]
    \setlength{\tabcolsep}{3pt}
    \centering
    \caption{3D RMS Error (less is better) in orbit propagation for different \ac{SRP} methods (rows) on different GPS satellite trajectories (columns), as well as their average (last column).
    Our average error is lower than any other baseline.
    Note, that while this confirms non-inferiority, our main aim is adding differentiability.}
    \label{tab:SimulationVerification}
    \begin{tabular}{lrrrrrrrrrrrrr}
\toprule
&\multicolumn{13}c{3D RMS error in meter}\\
\cmidrule{2-14}
&\multicolumn{12}c{Space Vehicle Number (SVN) \cite{enge1994global}}\\
\cmidrule{2-13}
\multicolumn1c{Method}&
\svn{63}&\svn{69}&\svn{67}&\svn{72}&\svn{68}&\svn{73}&\svn{65}&\svn{62}&\svn{71}&\svn{66}&\svn{64}&\svn{70}&Mean\\
\midrule
\method{Ball}&10.58&15.22&10.51&18.25&9.06&8.65&8.22&13.19&10.20&19.12&6.43&12.42&11.82\\
\method{ClassicGrid}&1.08&0.97&0.28&1.07&1.40&1.14&\winner{0.48}&0.88&\winner{0.39}&0.76&1.21&1.09&0.90\\
\method{OursGrid}&\winner{0.51}&\winner{0.47}&0.23&\winner{0.41}&0.44&\winner{0.58}&0.59&\winner{0.65}&0.67&0.60&\winner{0.50}&0.44&\winner{0.51}\\
\method{OursGridNB}&0.53&0.47&0.24&0.42&0.45&0.59&0.60&0.69&0.68&0.62&0.54&0.45&0.52\\
\method{OursNN}&0.61&0.56&\winner{0.20}&0.65&\winner{0.34}&0.63&0.52&0.72&0.55&\winner{0.32}&0.59&\winner{0.43}&0.51\\

\bottomrule
    \end{tabular}
\end{table*}

\paragraph{Discussion}
For qualitative inspection, \refFig{Forcemaps} shows force maps of four representative satellites.
When comparing the different force maps horizontally, we see that they depend on the geometry and materials.
We can see that adding bounces changes the forces when comparing the ``direct'' and ``indirect'' rows.
Both these aspects indicate that intricate simulation of solar pressure is required.

The effect of changing the design on the force is reported in \refFig{Design}.

\mycfigure{Forcemaps}{Force maps for different spacecraft geometry and materials.
In a force map, each pixel corresponds to one light direction in the spacecraft-fixed latitude-longitude (as defined in \cite{montenbruck2015gnss, bhattarai2019demonstrating}) and the color corresponds to the resulting 3D force, mapped from signed 3D to unsigned RGB.
For each, we show four different force maps (top to bottom):
First, the neural proxy used for design.
We use no design parameters for this visualization.
Second, the immediate simulator output.
Third and fourth, the direct, resp.\ indirect force, split and tone-mapped separately.
We see that force maps depend on geometry and that the neural proxy captures the function.
Convergence plots of the neural proxy are show in the neural row for each column, where blue is train and orange is test.
The ground truth of the NN is our simulation, while for simulation, the ground truth is unknown.}

\myfigure{Design}{Visualization of sampling a one-dimensional design space of changing solar array width (from left to right).
At a single pixel, the radiance is a random one out of many designs.
When integrating over a thousand samples, we see the variation in a Schr\"odinger-style image where all designs are present at the same time.
We train a neural network from that information (more precisely, from force, not radiance shown here for visualization) that can then be queried at random-access designs.
The third and fourth row show the positional and rotational force map of the minimal (left) and maximal (right) solar array width\revision{, with 3D force mapped to RGB as in \refFig{Forcemaps}}.
}

\begin{table}[]
    \centering
    \caption{Speed in samples per second (more is better) for different meshes (rows) in low (left) and high (right) geometric detail.}
    \label{tab:Speed}    \begin{tabular}{lrrrr}
\toprule
&
\multicolumn2c{Low}&
\multicolumn2c{High}\\
\cmidrule(lr){2-3}
\cmidrule(lr){4-5}
\multicolumn1c{Spacecraft}&
\multicolumn1c{Tri.}&
\multicolumn1c{Smp./s}&
\multicolumn1c{Tri.}&
\multicolumn1c{Smp./s}\\
\midrule
GRACE&
40&
307.1&
24,746&
37.7\\
CHAMP&
67&
315.0&
2,224&
7.1
\\
GPS&
140&
112.4&
4,548&
36.4\\
Swarm&
182&
87.1&
19,972&
31.7\\
\bottomrule
    \end{tabular}

\end{table}

A typical radiation map in a typical resolution image or a set of training samples (\ie 100\,k light directions and/or designs, with 1000 MC samples each) for a representative satellite model (\ie 1\,k faces) can be computed in less than 10 seconds on a Nvidia A100.
This is close to 100 million samples per second.

\mywfigure{Convergence}{0.55}{Convergence (see text).}

A detailed breakdown for different satellites is seen in \refTab{Speed}.
We use two sets of meshes: high resolution meshes as provided by the group at TU Delft for GRACE-FO, Swarm and CHAMP \cite{March2019_geometry}, and a second model for GPS2F, which we developed ourselves using openly accessible data about the spacecraft \cite{fisher1999gps,montenbruck2015gnss, IGSOrbits}.
These have unbalanced and degenerate meshing and are in the order of ten-thousands of polys.
We have re-created these meshes by box-modeling to only require hundreds of faces.
In both settings, millions of samples can be computed, reducing a fully detailed solar radiation pressure calculations to a matter of seconds.

\refFig{Convergence} shows the effect of the different variance reduction schemes we apply.
We study pure MC, MC with BRDF IS, MC with stratified sampling of the surface and both.
The vertical axis is MSE to a 10,000-sample reference in log scale. The horizontal axis is samples in linear scale. We see that variance is reduced by a factor of around hundred.

It could be asked if acceleration structures like BVH might be able to accelerate the simulation.
This might be true for a fixed design, but for sampling the product space of design and solar directions, the building of a a BVH will not pay off.
Every pixel is a design and every different design is a different BVH, so tracing is at least as expensive as building a new tree for every ray tree.
Even if the tree could simply be refit, refitting requires access to every primitive, which is exactly the loop our intersection already covers.
Our tracing is linear in the number of primitives while building is at least linear, but rather with an additional log factor.
\mywfigure{BVHComparison}{0.5}{BVH time and space (see text).}

Experimental evidence for this is seen in \refFig{BVHComparison}, where we implemented a linear BVH \cite{karras2013fast} also in JAX.
We vary the triangle count from a single to $2^{14}$; the largest design we use along the horizontal axis and measure compute time of a BVH (blue) and ours (red) in the left plot.
We upper-bound the time it takes to intersect a BVH by only building the BVH and not even implementing the intersection in JAX.
Even if time was available to build the BVH, it must be stored somewhere, resulting in memory to grow linear-time-log in triangle count as measured in \refFig{BVHComparison}, right.
Both these findings indicate that our trivial intersection is likely faster, and in terms of memory, the only feasible way to solve the task of intersecting many rays with many designs.

While we will differentiate by-proxy, in respect to all parameters, the radiation pressure simulation is differentiable already to some parameters by-construction.
Implemented in JAX, we can use autograd via \textsc{Grad} as long as we do not differentiate with respect to discontinuous parameters.
We have verified that we can discover the reflectance of the GPS-2F satellite, given its 3D geometry, and ground truth force map using gradient descent within seconds.
We will not rely on this property or use this option in this paper, though.

\mysubsection{Representation}{RepresentationVerification}
Next, we check if the NN is a close fit to the ground truth force.
To do this, we train the force proxy with a training data set and check two aspects: first, what is the relative error in percentage, averaged over all light directions?
Second, when using the proxy instead of a force map, do we get different results?

\begin{table}[]
    \centering
    \caption{Error of the NN proxy in percentage of force as well as the time required to saturate at \revision{3}\,\% difference to final test loss  (columns) for different satellites (rows).}
    \label{tab:NNError}
    \begin{tabular}{lrr}
\toprule
\multicolumn1c{Spacecraft}&
\multicolumn1c{Relative Error}&
\multicolumn1c{\revision{Time-to-3\%}}\\
\midrule
GRACE&3.6\,\%&13 s\\
CHAMP&4.5\,\%&11 s\\
GPS&2.6\,\%&13 s\\
Swarm&5.2\,\%&15 s\\
\midrule
Mean&4.0\,\%&13 s\\
\bottomrule
\end{tabular}

\end{table}

We find that the error in percentage is as low as 4\,\% on average, as seen in \refTab{NNError}.
Training a typical force map to the 3\,\% error, typically, requires 15 seconds on an NVidia A100 (see details in \refTab{NNError}).
We also find that this is faster to execute; A Nvidia A100 computes 630 million differentiable and noise-free \ac{SRP} samples per second.
Note, that a NN query always takes the same amount of time, hence we do not check how this varies across spacecrafts.

\refTab{Architectures} looks at architectural alternatives to confirm that two layers with 16 internal states form a good trade-off.
This experiment is performed using the important GPS-2F geometry, with a two-dimensional design space of reflectance and solar array length.

\begin{table}[]
    \centering
    \caption{Instrumentation of architectural alternatives (rows) according to parameter count (less is better), relative error (less is better), train time (less is better) and bandwidth (more is better).
    ``Relative'' means in percentage to our choice used in all other results, the first row.}
    \label{tab:Architectures}
    \begin{tabular}{c rrrr}
\toprule
\multicolumn1c{Architecture}&
\multicolumn1c{Parameters}&
\multicolumn1c{Err.}&
\multicolumn1c{Train}&
\multicolumn1c{BW}\\
\midrule
16$\times$16$\times$16&4,913&1.00$\times$&1.00$\times$&1.00$\times$\\
16$\times$16&289&1.04$\times$&0.87$\times$&1.62$\times$\\
8$\times$8&81&1.18$\times$&0.71$\times$&2.10$\times$\\
4$\times$4&25&1.60$\times$&0.71$\times$&2.57$\times$\\
\bottomrule
    \end{tabular}

\end{table}

\mysubsection{Optimization}{OptimizationVerification}
To determine whether the adjoint solver is fundamentally applicable to our problem domain, we consider a scenario where the initial conditions are known, the observed trajectories are available, and a reliable model of the forces is present—specifically, the GPS orbits introduced in \refSec{IntegratorVerification}.
We then set one aspect of this to be unknown (``known unknowns'') and use the adjoint solver to find it. In particular we consider the parameters listed in \refTab{Adjoint}: the Earth's mass, the Moon's mass, the Sun's mass and  --more advanced-- the Earth's J2 (The first zonal harmonic coefficient accounting for Earth’s oblateness in its gravity field model). We run each optimization 10 times and start from initial values $1\pm.5$ times the true value.

\begin{table}[h]
\centering
\caption{
Optimizing for known orbit unknowns: Every row is one known simulation parameter we pretend to not know and optimize for, as well as the true value, the value optimized, and the resulting relative as well as absolute error.}
\label{tab:Adjoint}
\begin{tabular}{l
S[table-format=1.2e1]
S[table-format=1.2e1]
S[table-format=1.2e1]
S[table-format=1.2e1]
}
\toprule
&
\multicolumn2c{Parameter}&
\multicolumn2c{Error}
\\
\cmidrule{2-3}
\cmidrule{4-5}
\multicolumn1c{Param.}&
\multicolumn1c{True}&
\multicolumn1c{Optim.}&
\multicolumn1c{Abs.}&
\multicolumn1c{Rel.}\\
\toprule
Earth&
5.97e24&
5.97e24&
1.93e10&
2.51e-14\\
Moon&
7.35e22&
7.35e22&
2.63e13&
3.21e-10\\
Sun&
1.99e30&
1.99e30&
6.86e21&
3.45e-9\\
J2&
1.08e-3&
1.08e-3&
2.09e-14&
1.93e-11\\
\bottomrule
\end{tabular}
\end{table}

\refTab{Adjoint} shows the outcome of this experiment.
Note that while the absolute errors might be large, they are small relative to the absolute scale of those parameters at hand.
This demonstrates that the adjoint approach is sufficient to identify known ground truth values when used in conjunction with our force model.

Solving the Earth mass problem to an accuracy  of 1\,\% from an initial Earth mass that is ten times larger takes less than four minutes on an Nvidia A100. 

\paragraph{Discussion}
One could consider using the noisy gradients that come out of the simulation directly without relying on a neural proxy.
We found this to diverge, and to be much slower to compute as solving complex ODE with noisy gradients is unstable.

When the orbit test for \texttt{GPS-06} is repeated after adding uniform noise of $10^{-10}$ / $10^{-5}$ / $10^{-2}$ the end-point errors are \qty{0.13}{\meter} / \qty{13.1}{\meter} / \qty{15}{\kilo\meter} off, where the error using our noise-free proxy is \qty{0.05}{\meter}.

Estimating noise down to $10^{-2}$ takes minutes for one single force map, while it it takes milliseconds for a proxy.
As no differentiable solar pressure solver exists, we can only speculate how fast or noisy it is, but a signal’s gradient typically is more noisy than the signal, making such an alternative even less attractive.
The noise from MC will also not be uniform, but worse, e.g., contain “fireflies”.
All this indicates using a proxy is more efficient and stable when using SRP in a inverse optimization approach.

We use 64 bit floating point, which is slower, but possible in JAX, as the smallest representable fp32 values are in the order of the effect of solar pressure itself!
For example \texttt{GPS-06} at 32 bit is off by \qty{272.3}{\meter} after \qty{12}{\hour}, while ours is off by \qty{0.05}{\meter} at 64 bit.

\mysection{Applications}{Applications}
As all the above works, we can now perform some actual designs and, even if we cannot deploy it to space, have presented a deck of validation examples to indicate it might work ``up there''.
An overview of the tasks we solve is shown in \refTab{Overview}.

In all the following experiments, the force model used includes the following forces: Earth gravity to degree and order 2, luni-solar gravity, and \ac{SRP}, eventually in combination with some other forces.
All applications are tested in \ac{MEO}, unless stated otherwise. 

  \newcommand{\app}[1]{\nameref{sec:#1}& \refSec{#1}}
  \begin{table*}[]
      \centering
      \caption{Overview of different applications for differentiable radiation pressure modeling.
      ``Target'' is a vector that is converted to a cost by the $L_2$ norm.}
      \label{tab:Overview}
      \begin{tabular}{lllllllllll}
          &&
          \multicolumn4c{Optimization}\\
          \cmidrule{3-6}
          \multicolumn1c{Task}&
          \multicolumn1c{Sec.}&
          \multicolumn1c{\rotatebox{90}{Refl.}}&
          \multicolumn1c{\rotatebox{90}{Shape}}&
          \multicolumn1c{\rotatebox{90}{Emm.}}&
          \multicolumn1c{\rotatebox{90}{Dyn.}}&
          \multicolumn1c{{Torq.}}&
          \multicolumn1c{Add. Forces}&
          \multicolumn1c{Design}&
          \multicolumn1c{Dims.}&
          \multicolumn1c{Target}\\
          \toprule
          \app{Approach}&
          \cmark&
          \xmark&
          \xmark&
          \xmark&
          \xmark&
          &
          Reflectance&
          $1$-D&
          Endpoint\\
          \app{AttitudeControl}&
          \xmark&
          \cmark&
          \xmark&
          \xmark&
          \cmark&
          &
          Three panel angles&
          $3$-D&
          Angular velocity
          \\
          \app{Chase}&
          \xmark&
          \cmark&
          \xmark&
          \xmark&
          \xmark&
          &
          Wing angle&
          $1$-D&
          Pos. rel. to client\\
          \app{CollisionAvoidance}&
          \cmark&
          \xmark&
          \xmark&
          \cmark&
          \cmark&
          &
          Reflectance&
          $6$-D&
          Coll. Avoid.\\
          \app{AtmosphereModelling}&
          \xmark&
          \xmark&
          \xmark&
          \xmark&
          \xmark&
          Atmo. drag&
          H\&P Atmo. params,&
          $39$-D&
          Mult. traject.\\
          \app{FormationFlight}&
          \cmark&
          \xmark&
          \xmark&
          \xmark&
          \cmark&
          &
          Six louver angles&
          $6$-D&
          Pos.+Rel. Vel
          \\
          \app{ShapeFromPressure}&
          \xmark&
          \cmark&
          \xmark&
          \xmark&
          \xmark&
          &
          Shape space&
          $7$-D&
          Trajectory
          \\
          \app{NeuralControl}&
          \xmark&
          \xmark&
          \xmark&
          \xmark&
          \cmark&
          &
          Policy NN&
          $5382$-D&
          Coll. Avoid. Rew.\\
          \app{OrbitCompute}&
          \xmark&
          \xmark&
          \xmark&
          \cmark&
          \xmark&
          Thermal&
          Six compute loads&
          $6$-D&
          Coll. Avoid.\\
          \bottomrule
      \end{tabular}
  \end{table*}

\mysubsection{Way-point intercept}{Approach}
\myfigure{PointApproach}{Point Approach task (\refSec{Approach}): spacecraft of different reflectance (line type) accumulate different offsets from a target  way-point (vertical) over time (horizontal).
When optimized (solid line) the spacecraft reaches the way-point (distance zero at time zero).}
\paragraph{Design}
We consider a \qty{600}{\kilo\gram} cube-shaped satellite with a surface area of \qty{50}{\meter\squared}, and ask what uniform reflectance it needs to reach a specified offset from its baseline trajectory.
The one-dimensional design space is the reflectance of all satellite faces.
While we do not claim this exact configuration is practical, spacecraft materials with a range of reflectance values are readily available \citep{sheldahl2015red,reyes2018characterization}.

\paragraph{Goal}
Starting from initial conditions taken from SVN 67 ephemeris \citep{enge1994global}, the goal is to reach a target way-point defined in the local orbital frame relative to the baseline trajectory.
We sample a random target offset on the order of meters yielding a 3D displacement of \qty{8.22}{\meter}.
Simulation spans \qty{12}{\hour} (\qty{2880}{steps} of \qty{15}{\second}).
No target velocity is specified; the spacecraft must pass through the way-point at the prescribed time with any velocity.
This experiment is intentionally minimal and serves as a baseline, with subsequent sections increasing both design dimensionality and terminal constraints (\eg docking in \refSec{FormationFlight}).

\paragraph{Result}
Our optimization finds a reflectance of 0.7 that will result in arriving at the desired point with an error of \qty{5}{\cm} (\refFig{PointApproach}).

\mysubsection{Attitude control}{AttitudeControl}

\paragraph{Design space}

A geometry with mass of \qty{30}{\kilo\gram} and a surface area of \qty{100}{\meter \squared} that comprises of a bus and three panels each laid along the body-fixed $x$,$y$,$z$-axes which can arbitrarily change their orientation around their longest axis (\refFig{AttitudeControl}).
There is non-zero initial angular velocity of \qtylist[list-units = single]{0.05;0.05;0.02}{deg/s} for $x$, $y$ and $z$ rotation, respectively.
Simulation is over six hours with a time step of \qty{30}{\second}.

\myfigure{AttitudeControl}{Attitude control task from \refSec{AttitudeControl}.
We consider a spacecraft in a circular orbit equipped with three solar arrays that can rotate about their longest axis.
Changing orientation over time (horizontal), results in rotational forces (different colors are different axes) that can we can simulate and differentiate, so as to modulate and dampen the overall angular velocity (vertical).
Our (solid lines) optimized result minimized angular velocity compared to random orientation (dotted).}

\paragraph{Goal}
We seek to dampen the existing rotational forces by moderating existing radiation pressure torques.
The panel angles over time are optimized to this end.
Hence, the loss is total scalar angular velocity (magnitude of 3D angular velocity).
This would typically be compensated for at the expense of precious thruster fuel.

\paragraph{Result}
The angular velocity over time is plotted in \refFig{AttitudeControl} with and without our optimization.
We see that without the control, the spacecraft maintains a high angular velocity.
In our solution, the initial rotation ebbs off.

\mysubsection{Chase}{Chase}

We now change the geometry of the spacecraft to fulfill a goal.

\paragraph{Design}

To this end, we consider a hypothetical spacecraft with flexible solar arrays as seen in \refFig{SciFi} which we scripted and equipped with materials (Black Kapton, Gold MLI, etc.) in Blender.
Recent research, such as \citet{banik2017roll}, demonstrates that flexible space structures, including solar arrays, may soon offer practical space engineering solutions. The spacecraft mass was set to \qty{1000}{\kilo\gram} and the total solar array area to \qty{100}{\meter\squared}.
The control parameters change the shape of the wing: they bend non-linearly in the ``coronal'' plane up to $\pm$\qty{60}{\degree} in both directions.
This strongly modulates the SRP.
\mywfigure{Chase}{0.45}{Chase (see text).}
\paragraph{Goal} We aimed to maneuver the chaser spacecraft to a target position within a defined window relative to a target satellite. The desired final offsets were \qty{0.75}{\meter} in the radial direction and \qty{12.5}{\meter} in the along-track. To achieve this, we optimized the solar array orientation over a period of 48 hours, with the objective of bringing the chaser closer to the target window through passive control. This optimization approach is applicable to typical rendezvous and proximity operations scenarios, where a chaser satellite must approach a client satellite.

\paragraph{Result} Through optimization of the solar array shape, the chaser spacecraft achieved a positional displacement of \qty{0.37}{\meter} in the radial direction, \qty{0.07}{\meter} in the cross-track, and \qty{6.29}{\meter} in the along-track over the 48-hour period—covering \qty{50}{\percent} of the required \qty{12.5}{\meter} transit via solar radiation pressure alone.

Without optimization, the total delta-v required to reach the target position would be \qty{1.17}{\milli\meter\per\second}. The optimization reduced this to \qty{0.58}{\milli\meter\per\second}, saving \qty{0.59}{\milli\meter\per\second}. This is a \qty{50}{\percent} reduction in propellant expenditure.
While a simple example, this demonstrates the potential of spacecraft shape optimization for reducing propellant use by leveraging the effects of SRP.

\myfigure{SciFi}{Radiation pressure maps for two hypothetical flexible satellite design: bird and flower.
The rendering show samples from the state, where the bird design bends its solar arrays along the $x$ axis, while the flower moves the solar arrays radially.
Below each rendering, we show the resulting force maps, sampled at different design parameters.
For the bird design, a cross is forming in the red area.
For the flower one, the green annulus is getting more pronounced.}

\mysubsection{Collision avoidance}{CollisionAvoidance}

\myfigure{CollisionAvoidance}{Collision avoidance task from \refSec{CollisionAvoidance}.
A satellite in a near-circular LEO \emph{(a)} can independently modulate the reflectance of each face over time \emph{(b)}.
The objective is to station-keep around a nominal orbit, except when approaching debris (hatched region), at which point it is required to create separation (vertical) by temporarily deviating from the nominal trajectory.
}

\paragraph{Design}
We consider a spacecraft that can dynamically change its reflectance (\refFig{CollisionAvoidance}, b).
A realized example of this idea is the IKAROS system where LCD panels were embedded into a solar sail \cite{mori2014overview}.
In our model, we use a cube geometry with mass of \qty{100}{\kg} and a surface area of \qty{50}{\meter\squared}. 
Simulation is over six days in \qty{120}{\second} steps.

\paragraph{Aim}
Produce a deviation in the nominal orbit of a spacecraft at a specific time at which a collision is likely to occur, and then return to the baseline orbit (\refFig{CollisionAvoidance}, a).
The loss is defined as
\begin{align}
\label{eq:CollisionLoss}
\loss(\trajectory) = 
\weightPost \, \deviationPost + 
\weightFinal \, \deviationFinal - 
\weightCol \, \deviationCol
\end{align}
where \deviationCol is the mean deviation from baseline near collision time, weighted by a Gaussian kernel ($\kernelWidth = \qty{3}{\hour}$) so that deviations closer to the collision time contribute more strongly; \deviationPost is the mean deviation after the maneuver; and \deviationFinal is the deviation at the final timestep.
The coefficients $(\weightPost, \weightFinal, \weightCol) = (2, 5, 17)$ were tuned empirically to balance collision avoidance and orbit recovery.
  
\paragraph{Result}
\refFig{CollisionAvoidance}, c) shows distance from the nominal trajectory over time.
Without optimization, the spacecraft passes close to the debris and subsequently drifts from its nominal orbit.
With optimized reflectance scheduling, the spacecraft creates sufficient separation at the collision time while returning to baseline afterward.

\mysubsection{Atmosphere modeling}{AtmosphereModelling}

\paragraph{Design}
Thermospheric density is one of the largest error sources in \ac{LEO} orbit prediction \citep{Mutschler2023CubeSatDensity}, and accurate density estimation from satellite constellation data is an active area of research in the satellite operations community \citep{Constant2025BringTheNoise}.

Atmospheric drag acceleration depends on local density $\rho$, spacecraft velocity $v$ relative to the co-rotating atmosphere, and the ballistic coefficient $C_D A/m$:

\begin{equation}
\mathbf{a}_\mathrm{drag} = -\frac{1}{2} \rho \, C_D \frac{A}{m} \, v^2 \, \hat{\mathbf{v}}
\end{equation}

where $C_D$ is the drag coefficient (a dimensionless quantity capturing the spacecraft's aerodynamic interaction with the rarefied flow), $A$ is the cross-sectional area, $m$ is the mass, and $\hat{\mathbf{v}}$ is the velocity unit vector \citep{montenbruck2000satellite}. Inverting observed trajectories for density requires accurate modeling of all other perturbations, particularly SRP, which acts at similar magnitudes in the upper thermosphere.

We simulate a constellation of 50 yaw-steering box-wing spacecraft (mass \qty{1633}{\kilo\gram}, cross-sectional area \qty{26.35}{\meter\squared}, $C_D = 2.2$) randomly distributed between \qty{350}{\kilo\meter} and \qty{720}{\kilo\meter} altitude with uniformly sampled right ascension of the ascending node and inclination. All orbits are circular. We use a constant drag coefficient for simplicity; our framework could compute attitude-dependent drag similarly to SRP, though we leave this to future work.

Truth trajectories are generated by propagating each spacecraft for one orbital period (${\sim}\qty{100}{\minute}$) at \qty{60}{\second} intervals through the \citet{harris1962time} density model with known parameters.

\myfigure{Atmosphere}{
Atmospheric density varies with altitude, latitude, and longitude \emph{(a)}.
We estimate density model parameters from satellite trajectories as described in \refSec{AtmosphereModelling}, visualizing traversal (blue) and sagittal (pink) slices through the spatial distribution.
Using our SRP model \emph{(b)} yields \qty{0.1}{\percent} mean absolute error relative to truth, while the cannonball baseline \emph{(c)} yields \qty{20.0}{\percent} error—demonstrating that SRP model fidelity directly impacts density retrieval accuracy.}

\paragraph{Aim}
Given only the observed trajectories and spacecraft geometry, recover the \citet{harris1962time} density lookup table through gradient-based optimization. This table specifies minimum and maximum density values at discrete altitudes, plus an exponent controlling angular variation between night-side and day-side density—39 parameters total for the altitude range spanned by our constellation. By comparing recovery using high-fidelity versus cannonball SRP models, we isolate the impact of SRP modeling fidelity on density retrieval.

\paragraph{Result}
\refFig{Atmosphere} shows the recovered density fields. With accurate SRP modeling, the optimizer correctly attributes trajectory perturbations to atmospheric drag, recovering density parameters to within \qty{0.1}{\percent} mean absolute error. The cannonball model conflates SRP and drag effects, yielding \qty{20.0}{\percent} error. Our results suggest that SRP model fidelity is a limiting factor for differentiable trajectory-based density estimation especially at higher altitudes.

In future work, this approach could complement neural density models \citep{perez2015neural} or be used to jointly invert for both density and space weather conditions.

\mysubsection{Formation flight}{FormationFlight}

\paragraph{Design}
In this task, we co-design the time-varying behavior of multiple spacecraft to perform formation flight by leveraging solar radiation pressure.
We consider two identical cube-shaped satellites, each with a mass of \qty{100}{\kilo\gram} and size \qtyproduct[product-units = single]{5.77 x 5.77 x 5.77}{\metre}, equipped with ``louver'' geometry on each of their six faces.
This geometry changes the apparent normal of each face by rotating 36 louvers as shown for three examples in \refFig{FormationFlight}, b.
All louvers on one face share the same orientation, giving six degrees of freedom per spacecraft.

\paragraph{Aim}
Both spacecraft start \qty{200}{\meter} apart in the along-track direction. The aim is for them to dock with one another below a certain relative velocity (\qty{\leq10}{mm/s}), leveraging only SRP from distinct louver configurations on each spacecraft (\refFig{FormationFlight}, a).
The simulation runs for \qty{32}{\hour} with a time step of \qty{60}{\second}.

\myfigure{FormationFlight}{Formation flight result from \refSec{FormationFlight}.
We optimize the time-varying louver geometry of two spacecraft to perform a controlled close approach \emph{(a)}.
Panel \emph{(c)} shows the relative motion in the along-track and height directions as defined from the target spacecraft; cross-track motion is negligible.
The design parameters are the louver orientations; three examples are shown in \emph{(b)}.
The aim is for the spacecraft to make contact (no attitude constraint) below a target relative velocity.
Random (dotted) louver orientations fail to achieve this, while the optimized configuration (solid) produces the desired behavior.
Panel \emph{(d)} shows the louver schedules: time on the horizontal axis, six faces on the vertical axis, one matrix per spacecraft.}

\paragraph{Result}
\refFig{FormationFlight}, c shows the relative position of spacecraft 2 with respect to spacecraft 1 in the radial/along-track plane.
The optimized trajectory first raises altitude to reduce relative velocity, then lowers altitude to re-accelerate toward the target, achieving rendezvous at the required realtive velocity.
The unoptimized case (louvers fixed shut) drifts around the target with excessive relative velocity.
\refFig{FormationFlight}, d shows the corresponding louver schedules.
The angles exhibit periods of constancy punctuated by rapid transitions, and are occasionally anti-correlated between spacecraft: tilting corresponding louvers in opposite directions modulates the relative acceleration between the two vehicles.
This demonstrates that SRP-based control can coordinate multiple spacecraft, not just a single one.

\mysubsection{Shape-from-pressure}{ShapeFromPressure}
\paragraph{Design}
Space object characterization typically relies on optical methods such as photometric light curves, which encode shape and reflectivity information in brightness variations over time \citep{linares2014shape,bradley2014lightcurve}.
Orbital dynamics, when considered, usually serve as a secondary constraint—for example, by providing attitude context for light curve interpretation.
We demonstrate the converse: given attitude knowledge, multiple correlated geometric parameters can be recovered from trajectory perturbations alone.
This assumption is reasonable for active spacecraft, which generally follow well-defined attitude laws (e.g., nadir-pointing, sun-tracking) \citep{montenbruck2015gnss}.

We consider a parametric family of satellite designs (\refFig{ShapeFromPressure}, a) in which bus width, height, and depth, panel width and height, and the reflectance of bus and panels can vary—seven parameters in total.
The parameter ranges are: \qtyproduct[product-units = single]{1.15 x 1.0 x 0.9}{\metre} to \qtyproduct[product-units = single]{4.60 x 4.0 x 3.73}{\metre} for the bus, \qtyproduct[product-units = single]{0.92 x 3.61}{\metre} to \qtyproduct[product-units = single]{3.68 x 14.45}{\metre} for the solar arrays, and the unit interval for reflectance.

\paragraph{Aim}
Recover the geometric parameters of a single satellite instance by fitting simulated radiation pressure perturbations to observed orbits (\refFig{ShapeFromPressure}, b) under known initial conditions.
We use 12 MEO orbits (approximately \qty{20000}{\kilo\meter} altitude) with varying inclination, eccentricity, and right ascension of the ascending node, each 6 hours in duration with \qty{60}{\second} time steps.
Each satellite follows a nadir-pointing attitude model, with one bus face oriented toward Earth and the solar arrays fixed within the orbit plane.
Observations are assumed noise-free.

\paragraph{Result}
\refFig{ShapeFromPressure}, c shows the evolution of estimated parameters (solid lines) toward the ground-truth values (dotted lines) from a random initialization.
Within a few iterations, all seven parameters converge: the system recovers satellite geometry from trajectory data and a model of how solar radiation pressure affects motion.

\myfigure{ShapeFromPressure}{Shape-from-pressure task as described in \refSec{ShapeFromPressure}:
Consider a parametric family of satellite designs \emph{(a)}, where each bar depicts a degree of freedom (panel width, bus height, etc.).
We optimize shape parameters to fit simulated radiation pressure perturbations to observed orbits—\emph{(b)} shows the trajectory data used—under known initial conditions.
\emph{(c)} shows parameter estimates (solid) converging to ground truth (dotted).}

This demonstrates that our approach handles high-dimensional design spaces well enough to solve the challenging inverse problem of inferring geometry from observed trajectories.
Practically, this capability could complement optical characterization; trajectory perturbations provide an independent geometric constraint.
Because spacecraft surface materials degrade in the space environment—altering optical and radiative properties over time \citep{miller2010degradation,engelhart2019spaceweathering}—trajectory-based inference could also provide a means to monitor material aging through its effect on radiation pressure response.
Joint estimation of attitude and geometry, and extension to noisy observations, are natural directions for future work.

\mysubsection{Neural control}{NeuralControl}
\paragraph{Design}
In previous tasks we have used our pipeline, which involves a neural proxy, to optimize interpretable physical parameters.
In this task, we instead use the pipeline to learn a neural policy \cite{grzeszczuk1998neuroanimator,izzo2024optimality}: the design space is now the parameters of a policy network that takes the world state as input and maps it to actions.
Such a policy is much faster to execute, requires no access to the future, and is better suited for deployment in a real mission.

\paragraph{Aim}
The goal is to obtain a policy that, based solely on state information (\eg sensor readings), selects the same actions that would be chosen by an optimization procedure.
The policy is trained using our simulation with access to privileged future information: at train time, the loss knows what the final reward for an action would be, while at test time, actions depend only on present information.
The network must therefore implicitly build an internal representation to predict the outcome and resulting reward.
For this task, we learn a policy for the collision avoidance scenario from \refSec{CollisionAvoidance}, using the same reward function.
The policy network is a three-layer MLP with 64 hidden units (5382 trainable parameters).
Inputs are: body-fixed sun direction (3D), spacecraft Euler angles (3D), angular velocity (3D), distance to desired orbit (1D), orbit phase (1D), and collision proximity (1D)---12 dimensions total. 
Outputs are the reflectance of each cube face (6D).

\paragraph{Result}
An ideal policy with limited information would achieve the same loss as full optimization with privileged information.
This lends itself to two measures of success: the ratio of loss between the neural policy and full optimization (ideally close to 1), and the ratio of execution time (ideally as high as possible).
We found the first ratio to be \qty{93}{\percent} $\pm$ \qty{11}{\percent} (averaged across 30 randomly selected initial conditions), and the second to be 72$\times$---losing \qty{7}{\percent} in performance for a 72-fold speedup.
This demonstrates that SRP-aware control policies can be distilled from our differentiable pipeline, showcasing their potential application to real-time autonomous maneuvers without onboard trajectory optimization.

\mysubsection{Compute in space}{OrbitCompute}

\myfigure{OrbitCompute}{Main result of \refSec{OrbitCompute}, where a \qty{500}{\kilo\gram} box-wing spacecraft is equipped with one GPU that emits heat \emph{(a)} on each of the six sides of the bus. The spacecraft surface area is \qty{80.85}{\meter\squared}.
This radiation results in deviation (vertical axis) from an orbit without compute; the dashed lines show this deviation for four arbitrary compute patterns \emph{(b)} varying over time (horizontal axis).
We optimize the allocation of compute over time \emph{(c)}, so as to create an orbit \emph{(d)} that avoids a space-time window (striped) in which the craft would collide with some space debris \emph{(e)}.
}

\paragraph{Design}
In this variant, we do not only consider radiation pressure from the sun, but also due to thermal radiation that is due to on-board compute.
The main reason to perform space-based compute is the high and time-constant solar power available \cite{glaser1968power}.
We would like to allocate compute on the spacecraft such that the effects of radiation pressure to compute and the sun are mitigated.
Similar optimizations are done on earth to distribute compute to use the cheapest, or most clean energy available \cite{xu2013temperature}.

We consider a cubic spacecraft with two panels that has compute on all six sides (\refFig{OrbitCompute}, a).
This leads to a 6D design space where each dimension scales the compute performed, and hence heat emitted, of one side.
Our simulation easily handles local sources in the same way it does handle distant sun light and the input to the neural proxy is now 6D radiation instead of 2D body-fixed sun directions. We model thermal emission as Lambertian following standard practice in spacecraft thermal recoil modelling \cite{turyshev2012pioneer}.

In particular, a single Nvidia H100 GPU requires \qty{700}{\watt} \cite{nvidia2026h100}, and hence a cluster of 100 GPUs produces \qty{70}{\kilo\watt}. For a Lambertian emitter, the momentum flux is reduced by a factor of 2/3 relative to a collimated beam \citet{turyshev2012pioneer}, yielding a recoil force of 
$
\qty{70}{\kilo\watt}
\times
2/3
\times
\qty{300000}{\km\per\second}
\approx
\qty{1.55e-4}{\newton}$.
The resulting acceleration \qty{ 1.55e-7}{\meter\per\second\squared} is in the same order of magnitude as solar pressure.

\paragraph{Aim}
The aim is to spatially balance compute inside the spacecraft to fulfill two goals.
First, that the resulting thermal radiation result in minimal deviation from a nominal trajectory.
Second, that a debris zone in time and space is avoided.
In other words, to ``jump'' over an obstacle as seen in \refFig{OrbitCompute}, c.
The compute is allocated such, that it results in heat dissipation of \qty{80000}{\watt} which maintains a temperature of \qty{100}{\celsius} for a spacecraft area of \qty{80}{\meter\squared}.

\paragraph{Result}
\refFig{OrbitCompute}, b shows the deviation from the nominal path.
When using arbitrary patterns of compute (not shown) this results in fluctuations around the nominal orbit (dotted lines).
Using spatial allocation of compute, this fluctuation can be controlled and used to avoid the obstacle (hatched areas) which the optimal solid black line avoids.
It does so without deviating more than required.
\refFig{OrbitCompute}, c shows the optimal compute pattern.

\mysection{Conclusion}{Conclusions}
In this work, we take an initial step toward predicting and systematically leveraging radiation pressure in the computational design of spacecraft and their behavior.
We make a simulator that, through greater parallelism and variance reduction, achieves higher accuracy and speed than current state-of-the-art models.
For the first time, we represent the force response to directional radiation for an entire family of satellites using a neural network.
This enables inverse tasks in space where a desired behavior yields a corresponding satellite design via AI-driven optimization.
Radiation pressure is typically the dominant non-conservative acceleration at altitudes where debris collision risk is highest \cite{schildknecht2007optical, mcknight2021identifying}. Our method can facilitate improved space situational awareness in these regions and thus contribute to mitigating debris proliferation.
Crucially, democratizing physics-based SRP computation has broad implications for space sustainability and traffic management.
This offers many new possible benefits to the precise orbit determination and space flight dynamics communities, where only a few groups \cite{kenneally2019faster, kenneally2020fast, ziebart2005combined, hladczuk2024grace} could previously make use such physics-based modeling of radiative forces.

\begin{acks}
We thank Yoaz Bar-Sever and the members of the NASA Jet Propulsion Laboratory Tracking Systems and Applications section for their feedback on early versions of this work.
Tobias thanks BS and IC for attitude control.    
\end{acks}

\bibliographystyle{ACM-Reference-Format}
\bibliography{paper}

\end{document}